\documentclass[doublecol]{epl2}
\usepackage{graphicx}
\usepackage{bm}
\usepackage{amssymb}
\usepackage{amsmath}
\newcommand{\be}{\begin{equation}}
\newcommand{\ee}{\end{equation}}
\newcommand{\beqn}{\begin{eqnarray}}
\newcommand{\eeqn}{\end{eqnarray}}

\title{Universal logarithmic terms in the entanglement entropy of $2d$, $3d$ and $4d$
random transverse-field Ising models}
\shorttitle{Universal logarithmic terms in the entanglement entropy\dots} 

\author{I. A. Kov\'acs\inst{1,2}\thanks{E-mail: \email{kovacs.istvan@wigner.mta.hu}} \and F. Igl\'oi\inst{2,3}\thanks{E-mail: \email{igloi.ferenc@wigner.mta.hu}}}

\shortauthor{I. A. Kov\'acs \etal}

\institute{
  \inst{1} Department of Physics, Lor\'and E\"otv\"os University, H-1117 Budapest,
P\'azm\'any P. s. 1/A, Hungary, EU\\
  \inst{2} Wigner Research Centre, Institute for Solid State Physics and Optics,
H-1525 Budapest, P.O.Box 49, Hungary, EU\\
  \inst{3} Institute of Theoretical Physics,
Szeged University, H-6720 Szeged, Hungary, EU
}

\date{\today}

\abstract{
The entanglement entropy of the random transverse-field Ising model is calculated by a numerical implementation
of the asymptotically exact strong
disorder renormalization group method in $2d$, $3d$ and $4d$ hypercubic lattices for different shapes
of the subregion. We find that the area law is always satisfied, but there
are analytic corrections due to $E$-dimensional edges ($1\leq E \leq d-2$).
More interesting is the contribution arising from corners, which is logarithmically divergent at
the critical point and its prefactor in a given dimension is universal, i.e. independent of the form of disorder.
}


\begin{document}
\maketitle

\section{Introduction}
To study the entanglement properties of quantum many body systems is a promising concept to understand
their topological and universal properties, in particular in the vicinity of a quantum phase-transition point\cite{entanglement_review,amico}.
Generally the entanglement between the subsystem, ${\cal A}$ and the rest of the system, ${\cal B}$,
in the ground state, $\left|\varPsi\right\rangle$
is quantified by the von Neumann entropy of the reduced density matrix,
$\rho_{\cal A}={\rm Tr}_{{\cal B}} | \varPsi \rangle \langle \varPsi |$ as:
${\cal S}=-{\rm Tr}_{\cal A}\left(\rho_{\cal A} \log_2{ \rho_{\cal A}}\right)$.
Generally ${\cal S}$ scales with the area of the interface separating ${\cal A}$ and ${\cal B}$.
In some cases, however, there are singular corrections to the area law.
In one-dimensional ($d=1$) systems ${\cal S}$ is logarithmically
divergent at a quantum critical point\cite{holzhey,vidal,Calabrese_Cardy04}: ${\cal S}=\frac{c}{3} \log_2 \ell+cst$. Here $\ell$ is the size of the subsystem
and the prefactor is universal, $c$ being the central charge of the conformal field theory. Recently one
considers also generalizations to R\'enyi entropy and the properties of the entanglement spectrum\cite{Calabrese_Lefevre}.

In higher dimensions our understanding about bipartite entanglement is far less complete, the known results
are almost exclusively about two-dimensional ($d=2$) models. Considering non-interacting systems, for
free bosons the area law\cite{area} is found to be satisfied even in gapless phases\cite{boson}. On the contrary, for gapless free-fermionic
systems with short-range hoppings and a finite Fermi surface there is a logarithmic factor to the area law\cite{fermion}.
In interacting $d=2$ systems the area law is generally found to be satisfied, but in gapless phases and in quantum
critical points there are additional logarithmic terms, which are expected to be universal. This has been
demonstrated for the $d=2$ transverse-field Ising model \cite{2d_Ising} and for the antiferromagnetic Heisenberg model\cite{2d_Heisenberg}. For the latter the logarithmic terms are associated
to two sources: i) corners on the boundary of the subsystem and ii) non-trivial topology in the bulk.
There is another class of $d=2$ critical systems described by $d=2$ conformal field theory,
the prototype being the square lattice quantum dimer model\cite{rokhsar_kivelson}. For these models non-perturbative
analytical and numerical results are available and the log-correction to the area law is shown to be
universal and related to corners\cite{2d_conf}.

Besides pure systems there are also investigations about the entanglement properties of quantum models
in the presence of quenched disorder\cite{refael}. In $d=1$ random systems (random antiferromagnetic Heisenberg and XX models,
random transverse-field Ising model (RTIM), etc.) the critical point is controlled by a so called
infinite disorder fixed point (IDFP)\cite{danielreview}, which can be conveniently
studied by the strong disorder renormalization group (SDRG) method\cite{mdh,im}.
Using this approach logarithmic entanglement entropy is found with a universal
prefactor\cite{refael_moore04}, which has been numerically checked by density-matrix
renormalization\cite{laflorencie} and by free-fermionic methods\cite{igloi_lin08}.
For ladders of the RTIM the same scaling behavior of the entropy is found\cite{ladder} as in $d=1$. In $d=2$ the entanglement entropy of
the RTIM has been studied by the SDRG method in two papers with conflicting results
at the critical point.
Lin \textit{et al}\cite{lin07} has used periodic systems up to linear size $L=64$ and the numerical
results are interpreted in terms of a double-logarithmic factor to the area law: ${\cal S} \sim \ell\ln\ln \ell$.
In a subsequent study Yu \textit{et al}\cite{yu07} has used open systems up to $L=160$
and the numerical data are fitted with a logarithmic correction to the area law: ${\cal S}=a\ell+b \ln \ell$.
This type of choice of the singularity is motivated by the similar form of the entropy in $d=2$
conformally invariant models\cite{2d_conf}, although the logarithmic correction is not attributed to corner effects but to percolation
of correlated clusters.

In the present work we revisit the problem of scaling of the entanglement entropy in the $d=2$
RTIM and use an improved numerical algorithm of the SDRG method\cite{2dRG,ddRG}. We have studied finite systems up to
$L=2048$ and also investigated the entropy of the same model in $d=3$ and $d=4$. Our
goal is to answer the following basic questions.\begin{enumerate}
                                                       \item How the criticality of the RTIM is manifested in the
singular behavior of the entanglement entropy?
\item What is the physical origin of this singularity, corner and/or bulk effects?
\item Is this singularity universal and independent of the form of disorder?
\item Is it related to the diverging correlation length?
                                                      \end{enumerate}
The rest of the paper is organized as follows. After definition of the RTIM we recapitulate the basic steps
of the SDRG method to calculate the entanglement entropy. Our results
for $d=2$, $3$ and $4$ are presented in more details at the critical point and afterwards outside the critical point.
Our Letter is closed by a Discussion and a detailed derivation is presented in the Appendix.

\section{Model and the SDRG method}

The RTIM is defined by the Hamiltonian:
\be
{\cal H} =
-\sum_{\langle ij \rangle} J_{ij}\sigma_i^x \sigma_{j}^x-\sum_{i} h_i \sigma_i^z\;,
\label{eq:H}
\ee
in terms of the $\sigma_i^{x,z}$ Pauli-matrices at sites $i$ (or $j$) of a hypercubic lattice.
The nearest neighbor couplings, $J_{ij}$, and the transverse fields, $h_i$,  are independent random numbers,
which are taken from the distributions, $p(J)$ and $q(h)$, respectively. Following Refs.\cite{2dRG,ddRG} we
have used two disorder distributions, for both the couplings are uniformly distributed in $\left[ 0,1\right]$.
For \textit{box-$h$} disorder the distribution of the transverse-fields is uniform in $\left[ 0,h_b\right]$, whereas for the
\textit{fixed-$h$} model we have $h_i=h_f,~\forall i$.
The quantum control parameter is defined as $\theta=\ln(h_b)$ and $\theta=\ln(h_f)$, respectively.

\begin{figure}[!ht]
\begin{center}
\onefigure[width=3.2in,angle=0]{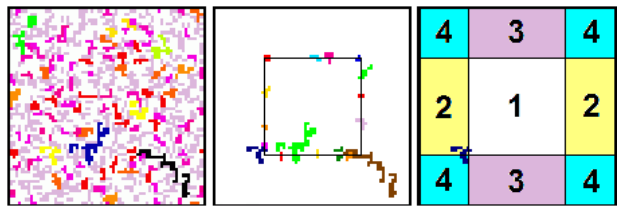}
\end{center}
\vskip -.5cm
\caption{
\label{fig_1} (Color online) Renormalized spin clusters (left panel) and those connected clusters,
which have a contribution to the entanglement
entropy (middle panel) in a $64 \times 64$ system with a subsystem of $32 \times 32$ spins (fixed-$h$ disorder at the
critical point).
(In the left panel clusters with the same mass are represented by the same color (greyscale), in the
middle panel spins in the same cluster are denoted in this way.) Right panel: partition of the sample into four
squares, denoted by $1,2,3$ and $4$ and for four slabs, each being composed of two
squares as: $(1,3);(2,4);(1,2)$ and $(3,4)$. In this sample and with this subdivision there is one ``corner''-cluster
(see at the left-low corner of $1$), which contributes to all the four slab subsystems, but does not contribute to all the four square subsystems, see text.
Note that periodic b.c.-s are used and the average value of connected and corner clusters in this geometry are measured $18.7$ and
$0.22$, respectively.}
\end{figure}

The ground state of the RTIM is calculated by an improved numerical algorithm of the SDRG method\cite{2dRG,ddRG}. During the SDRG\cite{im} the largest local terms in the Hamiltonian in Eq.(\ref{eq:H}) are successively eliminated and new Hamiltonians are generated through perturbation calculation. For the
RTIM two types of decimation are performed. i) If the decimated
term is a strong coupling, say $J_{ij}$,  then the two sites, $i$ and $j$ are \textit{aggregated} to a new
effective spin cluster. ii) If the decimated term is a strong transverse field, say $h_i$, then
the site, $i$ is \textit{eliminated}. After decimating all
degrees of freedom the ground state of the system is found as a collection of independent ferromagnetic clusters of
various sizes; each cluster being in a GHZ state\cite{refael_moore04,lin07,yu07}: $\frac{1}{\sqrt{2}}\left(|\uparrow \uparrow  \dots  \uparrow\rangle +
|\downarrow \downarrow \dots  \downarrow\rangle \right)$. Each cluster contributes by an amount of $\log_2 2=1$ to the
entanglement entropy if it is shared by the subsystems, otherwise the contribution is $0$.
Thus calculation of the entanglement entropy for the RTIM is equivalent to a cluster counting problem. This is
illustrated for $d=2$ in Fig. \ref{fig_1} (left and middle panels).
The structure of the GHZ clusters is different in the ferromagnetic phase, $\theta<\theta_c$, when there is a giant
cluster and in the paramagnetic phase, $\theta>\theta_c$, when all clusters have a finite extent.
The location of the critical point, $\theta_c$, has been calculated previously for the two random distributions
in $d=2$ Ref.\cite{2dRG} and in $d=3$ and $4$ in Ref.\cite{ddRG}.

\section{Results at the critical point}

We have calculated the entanglement entropy of the RTIM at the critical point in finite hypercubic samples of linear size, $L$
with full periodic boundary conditions (b.c.-s),
the largest sizes for box-$h$ (fixed-$h$) disorder being $2048~(1024)$, $128~(64)$ and $48~(24)$ for $d=2$, $3$ and $4$, respectively. (In the latter case the clusters are more compact, contain more sites and thus the analysis
of the entropy is more involved.) We have considered $d$ different
geometries, in which in $0 \le D \le d-1$ directions ${\cal A}$ extends to the full length of the system, $L$ and has periodic b.c.-s, whereas in the other directions its length is $\ell<L$. The three possible geometries for $d=3$ are illustrated in
the inset of Fig. \ref{fig_2}. Only in the cube geometry with $D=0$ there are corners, whereas for $D=d-1$ in the slab geometry
the interface contains no edges. For a given random sample and for each geometry we have averaged the entanglement entropy for every possible position (and orientation) of ${\cal A}$ and subsequently we have averaged over several samples. The
typical value of realizations being $40000$ but even for the largest sizes we had at least $10000$ samples.
For a given realization the extra computational time needed to perform
the cluster counting problem for the entropy is ${\cal O}(L^{2(d-D)})$, which can be speeded
up in the slab geometry, see next subsection.

\subsection{Slab geometry}

We start our investigations in the slab geometry, 
where the entanglement entropy of a sample averaged over all positions
can be written in a simple closed form in terms of cluster statistics. Here we announce the result, details
of its derivation can be found in Ref.\cite{perco} In this algorithm we consider that axis, say the $z$-axis,
which is perpendicular to the surface of the slab and we measure the $z$-coordinate of the points of the clusters.
For each cluster we arrange the different $z$ values as $z_1<z_2< \dots <z_k$ and
define the difference between consecutive $z$-values,
$h_i={\rm min}[z_{i+1}-z_{i},L-(z_{i+1}-z_{i})]$, $i=1,2,\dots k-1$;
$h_k={\rm min}[z_{k}-z_{1},L-(z_{k}-z_{1})]$ . Repeating this measurement for
all clusters we calculate the statistics of the $h_i$ differences:
$n(j)$ being the number of distances with $j=h$. The position
averaged entanglement entropy of the sample is then given for $\ell \leq L/2$:
\be
{\cal S}_{\rm slab}^{(d)}(L,\ell)=\dfrac{1}{L}\sum_{i=1}^{\ell}\sum_{j=i}^{L/2} n(j)\;.
\ee
This type of algorithm
works in ${\cal O}(L)$ times, which is to be compared with the performance of the direct cluster counting
approach: ${\cal O}(L^2)$.

\begin{figure}[!ht]
\begin{center}
\onefigure[width=3.2in,angle=0]{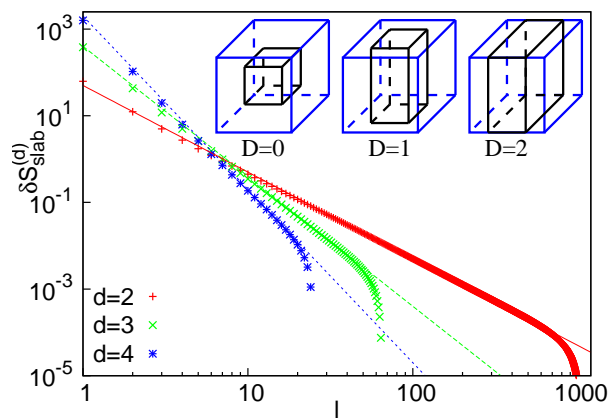}
\end{center}
\vskip -1cm
\caption{
\label{fig_2} (Color online) Finite difference of the entropy in the slab geometry,
$\delta {\cal S}_{\rm slab}^{(d)}(L,\ell)$, as a function of the width $\ell$
for $d=2,3$ and $4$ for the largest sizes, $L$, for box-$h$ disorder. (The error is smaller
than the size of the symbols.)
The asymptotic forms, $\sim \ell^{-d}$, are indicated by straight lines.
Inset: Three geometries of the subsystem ${\cal A}$ studied in $d=3$: cube ($D=0$), column ($D=1$) and slab
($D=2$).}
\end{figure}

In the slab geometry, the area is independent of $\ell$ and any singular contribution to the area
law can only be of bulk origin. 
In our study we have fixed $L$ to its largest value and calculated the entropy per area,
$a_{d-1}(L,\ell)$ for varying $\ell$. For $1 \ll \ell \ll L$ we have found that $a_{d-1}(L,\ell)$ approaches
a constant with a correction term: $\sim \ell^{-d+1}$.
To illustrate this relation we have calculated the finite difference:
$\delta {\cal S}_{\rm slab}^{(d)}(L,\ell)={\cal S}_{\rm slab}^{(d)}(L,\ell+1)-{\cal S}_{\rm slab}^{(d)}(L,\ell)$ as a function of $\ell$,
which has the behavior: $\delta {\cal S}_{\rm slab}^{(d)}(L,\ell) \sim \ell^{-d}$ as shown in Fig.\ref{fig_2} for $d=2,3$
and $4$. This type
of \textit{non-singular} contribution to the entropy in the slab geometry can be interpreted
in the following way. Due to the
finite width of the slab only those correlated domains can effectively contribute to the entropy, which have a finite extent
$\xi \lesssim \ell$. (Much larger clusters have typically no sites inside the slab.) 
Finite-size corrections are due to clusters with $\xi \approx \ell$, the number of these blobs
scales as $n_{bl} \sim (L/\ell)^{d-1}$ and each has the same correction to the entropy, which then
scales as $\sim n_{bl}$ in agreement with the scaling Ansatz and with the numerical data in Fig.\ref{fig_2}.
In the SDRG reasoning used in Ref.\cite{lin07} and which has lead to a $\ln\ln \ell$ multiplicative
correction one assumes the existence of several ($\ell$-dependent number of) independent large clusters in a $\xi \approx \ell$ blob,
which is in contradiction with the results of the present large-scale calculation.

\subsection{Column geometry}
\label{sec:column}

In the column geometry, see the middle panel in the inset of Fig. \ref{fig_2}, there are corrections to the area law due to edges.
Let us consider an $E$-dimensional edge ($1 \le E < d-1$) with a total surface, $f_E \sim L^{E}$, so that its contribution to the entropy is given by: $a_E f_E$. We have found,
that the $a_E$ prefactors have alternating signs: $a_{E}/a_{E-1} <0$ and $a_{d-2} <0$. Here using the same reasoning as in the slab geometry the correction to the edge contribution per surface
is given by: $a_E-a_E(\ell) \sim \ell^{-E}$, which result has been checked numerically. Thus
we can conclude that the contributions to the entropy due to edges are also \textit{non-singular}
and singular contributions can only be obtained at corners.

\subsection{Cube geometry}
\label{sec:cube}

\begin{figure}[!ht]
\begin{center}
\includegraphics[width=3.2in,angle=0]{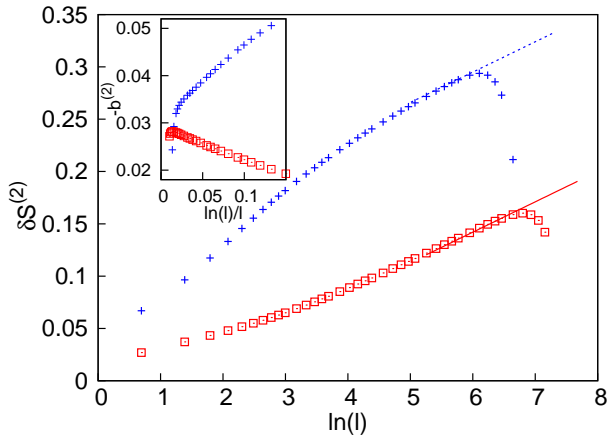}
\end{center}
\vskip -0.5cm
\caption{
\label{fig_3} (Color online) Corrections to the area law in $d=2$ as a function of $\ln \ell$ for fixed-$h$ ($+$, $L=1024$) and box-$h$ ($\Box$, $L=2048$) disorders. The asymptotic behaviors for large $\ell$ are indicated by straight lines having
the same slopes: $b^{(2)}=-0.029$. Inset: effective, $\ell$-dependent $-b^{(2)}$ parameters obtained from two-point fits.}
\end{figure}

In order to check the corner contributions to the entropy, ${\cal S}_{\rm cr}^{(d)}(\ell)$,
we study here cube subsystems, as shown in the right panel of the inset of Fig. \ref{fig_2}. In this
case we write ${\cal S}_{\rm cube}^{(d)}(\ell)$ in the general case as:
\be
{\cal S}_{\rm cube}^{(d)}(\ell)=a_{d-1} f_{d-1} + \sum_{E=1}^{d-2} a_E f_{E} + {\cal S}_{\rm cr}^{(d)}(\ell)\;.
\label{S^d}
\ee
where the corner contribution has the sign: $(-1)^{d+1}$, which is opposite to the sign of $a_1$.
In $d=2$, when the subsystem is a square, the second term in Eq.(\ref{S^d}) is missing and we obtain
accurate estimates for the corner contribution by evaluating the difference: $\delta {\cal S}^{(2)}(\ell)\equiv{\cal S}^{(2)}(\ell)-2{\cal S}^{(2)}(\ell/2) \approx {\cal S}_{\rm cr}^{(2)}(\ell)-2{\cal S}_{\rm cr}^{(2)}(\ell/2)$.
This is presented in Fig.\ref{fig_3} as a function of $\ln \ell$ for the two types of disorder using the largest finite systems.

For large $\ell$ the data approach a linear logarithmic dependence, $\delta {\cal S}^{(2)}(\ell) \simeq {\cal S}_{\rm cr}^{(2)}(\ell) + cst \simeq -b^{(2)} \ln \ell + cst$.
We have calculated effective, $\ell$-dependent
$b^{(2)}$ values from two point fits, which are presented in the inset of Fig.\ref{fig_3}. From their extrapolation we obtain
the estimate, $b^{(2)}=-0.029(1)$, for both types of disorder, which is to be compared with
$b^{(2)}=-0.019(5)$ by Yu \textit{et al}\cite{yu07} calculated in a much smaller system with box-$h$ disorder.

In higher dimensions, $d \ge 3$, the corner contribution represents only a very small fraction of the
entanglement entropy and thus its estimate through a direct analysis of Eq.(\ref{S^d}) contains rather large errors.
We can, however, circumvent this problem by considering samples with $\ell=L/2$, when ${\cal S}_{\rm cr}^{(d)}(\ell)$
is expressed as appropriate combination of the entropies of subsystems with different shapes for $D=0,1,\dots,d-1$.
This calculation is presented in the Appendix and
illustrated in the right panel of Fig.\ref{fig_1} for $d=2$. Here a given $L \times L$ sample contains four
square subsystems and also four slab subsystems. In the two geometries the accumulated boundary between
the subsystems and the environment is the same, thus the difference between the accumulated
entropies gives the corner contribution: ${\cal S}^{(2)}_{\rm cr}={\cal S}^{(2)}_{\rm square}-{\cal S}^{(2)}_{\rm slab}$.
This contribution is not zero, since the so called ``corner'' clusters, which have no sites in one or two non-contacted
$90$-degree corners (see in the right panel of Fig.\ref{fig_1}),
provide different contributions to ${\cal S}^{(2)}_{\rm square}$, than to ${\cal S}^{(2)}_{\rm slab}$.

\begin{figure}[!h]
\begin{center}
\onefigure[width=3.2in,angle=0]{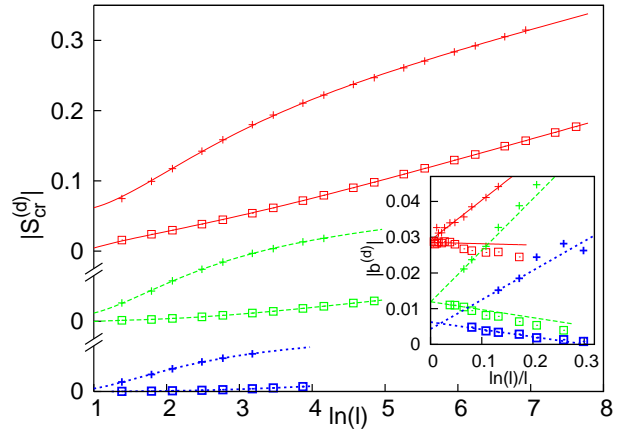}
\end{center}
\vskip -1cm
\caption{
\label{fig_4} (Color online) Corner contribution to the entanglement entropy as a function of
$\ln \ell$, for $d=2,3$ and
$4$ from up to down for fixed-$h$ ($+$) and box-$h$ ($\Box$) disorders. Note that the position of $0$ of the vertical axis
is shifted by $0.1$ with $d$. In the
inset the effective prefactors of the logarithm are shown as calculated by two-point fit.
The extrapolated values are disorder independent: $b^{(2)}=-0.029(1)$, $b^{(3)}=0.012(2)$ and $b^{(4)}=-0.006(2)$. 
The error of data is smaller than the size of the symbols in the main panel and these are smaller than
twice of the size of the symbols in the inset. The lines through the points are guide to the eye.}
\end{figure}

The corner contribution to the entanglement entropy at the critical point with $\ell=L/2$
for different values of $L$ are presented in Fig. \ref{fig_4} for $d=2,3$ and $4$
and for the two disorder distributions. For $d=2$ the variation with $\ln L$ is similar to that
obtained by the direct analysis in Fig.\ref{fig_3}. Also for $d>2$ asymptotically a logarithmic
increase is found ${\cal S}_{\rm cr}^{(d)}(\ell=L/2) \simeq b^{(d)} \ln \ell + cst.$ and the prefactors
$b^{(d)}$ are estimated by two-point fits. These are shown in the inset of Fig.\ref{fig_4}. 
Their extrapolated values are found to be disorder independent, thus universal and these are listed in the caption
of Fig. \ref{fig_4}.
For $d=2$ this coincides with the value calculated previously in Fig.\ref{fig_3}.

The $\ln \ell$ dependence of the corner contributions to the entropy can be understood with the example
of two-point clusters. (Here we remark that during renormalization spins are glued together to form new
effective spin variables and a final spin cluster, which appears in the right panel of Fig.\ref{fig_1} is also the result
of the aggregation of two effective spins.) It is easy to see, that
in $d$ dimension a two-point cluster is a ``corner'' cluster if the two points are located
in two such hypercubes, which are connected by the main diagonal. If the relative coordinates of the two-point cluster
are $0 \le x_j \le L/2,~j=1,2,\dots,d$ (due to periodic b.c.-s) then its accumulated contribution to the corner-entropy
(obtained by averaging over all possible positions) is
$-2 \prod_{j=1}^d(-x_j/L)$. The probability of having a two-site
cluster of a length, $r$, is given by the average pair correlation function, $C_{\rm av}(r)\approx n_r^2$,
where $n_r \sim r^{-d}$ is the density of non-decimated sites (when the typical length between existing effective
spins is $r$).
The average contribution to the corner-entropy can be estimated as:
${\cal S}_{\rm cr}^{(d)}(\ell) \sim - \int_1^{\ell} {\rm d} x_1 \dots \int_1^{\ell} {\rm d} x_d \prod_{j=1}^d(-x_j/r^2)
\sim (-1)^{d+1} \int_1^{\ell} (r^{d-1} r^d)/r^{2d} {\rm d} r \sim (-1)^{d+1} \ln \ell$, which is logarithmically divergent
in any dimension, in agreement with the numerical results in Fig. \ref{fig_4}.
\section{Results outside the critical point}
We have also studied the behavior of the corner-entropy outside the critical point and measured
${\cal S}_{\rm cr}^{(d)}(L,\delta)$ as a function of $\delta=\theta-\theta_c$.
In the ordered phase, $\delta<0$, and for $\xi<\ell$ the giant cluster
behaves as a so called global cluster, which has $1$ contribution to the entropy for all position, orientation and shape of the subsystem. As shown in the Appendix in odd (even) dimensions, after averaging for all positions a global cluster has a contribution $2^{1-d}$ ($0$) to the corner entropy. Approaching the critical point for $\xi \gtrsim \ell$ these giant
clusters have a finite, but $\delta$-dependent contribution, that we omit in the following analysis.

\begin{figure}[!ht]
\begin{center}
\onefigure[width=3.2in,angle=0]{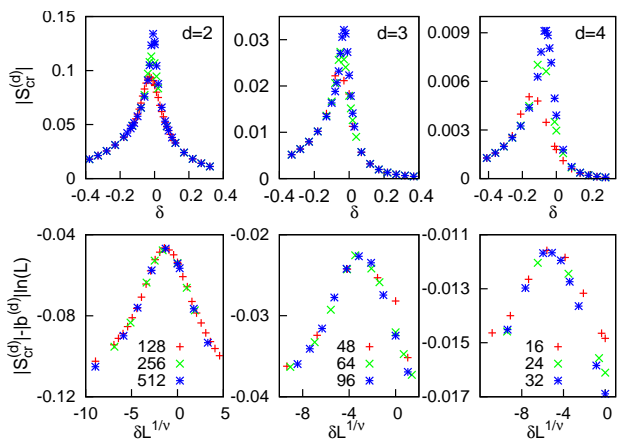}
\end{center}
\vskip -1cm
\caption{
\label{fig_5} (Color online) Upper panels: entropy contribution of the corners ${\cal S}_{\rm cr}^{(d)}(L,\delta)$ in $d=2,3$ and $4$ for different system sizes as the function of the control parameter for box-$h$ disorder. In
the lower panels scaling collapse of the central part of ${\cal S}_{\rm cr}^{(d)}(L,\delta)$ is shown, see text.}
\end{figure}

In the upper panels of Fig. \ref{fig_5} ${\cal S}_{\rm cr}^{(d)}(L,\delta)$ is presented
as a function of $\delta$ for different finite systems for box-$h$ disorder.
For any $d$ the corner-entropy is extremal around the critical point and its value
outside the critical point is well described with the substitution: $\ell \to \xi$, with $\xi \sim |\delta|^{-\nu}$
being the correlation length. Close to the critical point it satisfies the scaling relation:
$-{\cal S}_{\rm cr}^{(d)}(L,\delta)+b^{(d)} \ln L = f(\delta L^{1/\nu})$, as illustrated
in the lower panels of Fig. \ref{fig_5}. Here we have used our previous estimates for the
correlation length critical exponents\cite{2dRG,ddRG}, $\nu=1.24,~0.98$ and $0.78$
in $d=2,3$ and $4$, respectively.

\section{Conclusion}
We have studied the entanglement entropy of the RTIM in the vicinity of the quantum critical point
in dimensions $d=2,3$ and $4$ by an efficient numerical implementation of the SDRG method. Since the critical
properties of the RTIM are governed by IDFP-s\cite{2dRG,ddRG,MG12} at which the SDRG becomes asymptotically exact also our results
about the singularities of the entropy tend to be exact for large scales.
We expect that our finite-size results are already in the asymptotic regime, which is supported by the fact that
the singularity parameters obtained are disorder independent.
We have demonstrated that the area law is
satisfied for $d \ge 2$ and there is a singular correction to it in the form of $b^{(d)} \ln \ell$. This
correction is shown to be attributed to corners, related to the
diverging correlation length and universal, i.e. disorder independent.

Our investigations can be extended and generalized to several directions. First we mention that the extremal
behavior of the corner entropy at the critical point makes a possibility to detect and define sample
dependent critical points, a concept which has already been applied in $d=1$\cite{ilmr08}. It is of interest to
study the possible singularities of the entropy per area and the edge contributions per surface
as a function of $\delta$ around the critical point for $d \ge 2$. One can also study the entanglement
properties of diluted transverse-field Ising models for $d \ge 2$ having critical properties related to classical percolation\cite{senthil_sachdev}. Finally, we mention dynamical aspects
of the entanglement entropy after a sudden change of the parameters in the Hamiltonian at time $t=0$.
This question has been recently studied\cite{dynamics} in $d=1$ and an ultraslow increase of the entropy is found:
${\cal S}(t) \sim \ln \ln t$, if the quench is performed to the critical point of the system. For $d \ge 2$
one expects a similar time-dependence of the corner-contribution: ${\cal S}_{\rm cr}^{(d)}(t) \sim \ln \ln t$.

\acknowledgments
This work has been supported by the Hungarian National Research Fund under grant No OTKA K75324 and K77629 and by a German-Hungarian exchange program (DFG-MTA).
We thank to P. Sz\'epfalusy and H. Rieger for useful discussions.


\section{Appendix: Corner-contribution to the entropy for $\ell=L/2$}
Here we show how the corner-contribution to the entropy for a given sample can be deduced from the entropies measured
in different shapes of the subsystems.

We consider a $d$-dimensional hypercubic system with linear length $L$ with full periodic boundary conditions. Inside the hypercubic system we select subsystems of different shapes, which span the system in $D=0,1\dots ,d-1$ directions, but restricted to length $L/2$ in the others. (See the inset of Fig.\ref{fig_2}.) The so defined subsystems have hyperfaces of dimension $\Delta=D,D+1,\dots ,d-1$, and the surface of a $\Delta$-dimensional unit is
$\varrho_D(\Delta)=L^D2^{D-\Delta}$, while the number of equivalent hyperface units is given by
\be
n^{(d)}_D(\Delta)=2^{d-\Delta}\binom{d-D}{\Delta-D}\;.
\ee
We measure the entanglement entropy in this system, ${\cal S}^{(d)}_D$, for all different $D$, averaged over the $L^{d-D}$ possible positions and over the $\binom{d}{D}$ orientations of the subsystem. The entanglement entropy is written as the sum of the contributions of the different dimensional hyperfaces:
\be
{\cal S}^{(d)}_D=\sum_{\Delta=D}^{d-1}{\cal S}^{(d)}_{D}(\Delta)\;,
\ee
which is to be compared with ${\cal S}^{(d)}_0 \equiv {\cal S}^{(d)}_{\rm cube}$ in Eq.(\ref{S^d}).
First we note that
\be
\frac{{\cal S}^{(d)}_{D}(\Delta)}{{\cal S}^{(d)}_{0}(\Delta)}=\frac{\varrho_D(\Delta)n^{(d)}_D(\Delta)}{\varrho_0(\Delta)n^{(d)}_0(\Delta)}=2^D\dfrac{\binom{d-D}{\Delta-D}}{\binom{d}{\Delta}}\;,
\ee
where $\varrho_D(\Delta)n^{(d)}_D(\Delta)$ and $\varrho_0(\Delta)n^{(d)}_0(\Delta)$ are the total areas of the given
hyperface measured in the two shapes.
In this way we obtain:
\be
{\cal S}^{(d)}_D=\sum_{\Delta=D}^{d-1}2^D\dfrac{\binom{d-D}{\Delta-D}}{\binom{d}{\Delta}}{\cal S}^{(d)}_{0}(\Delta)\;.
\ee
It is straightforward to check, that this expression can be inverted to obtain the entropy contribution in the cube geometry:
\be
{\cal S}^{(d)}_0(\Delta)=\sum_{D=\Delta}^{d-1}{\frac{(-1)^{D-\Delta}}{2^{D}}\binom{d}{D}\binom{D}{\Delta}{\cal S}^{(d)}_{D}}\;.
\ee
As a special case for ${\cal S}^{(d)}_0(0)\equiv {\cal S}_{\rm cr}^{(d)}$ we obtain for the corner contribution:
\be
{\cal S}^{(d)}_0(0)=\sum_{D=0}^{d-1}{\left(-\frac{1}{2}\right)^{D}\binom{d}{D}{\cal S}^{(d)}_{D}}\;.
\label{corners}
\ee
As an application, we calculate the contribution of a global cluster to ${\cal S}^{(d)}_0(0)$. A cluster is global by our definition, if its entropy contribution is $1$ for all positions, orientations and shapes ($D$) of the subsystems, thus
\begin{eqnarray}
&\nonumber {\cal S}^{(d)}_0(0)_{({\rm global})}=\sum_{D=0}^{d-1}(-2)^{-D}\binom{d}{D}=\\ &\dfrac{1-(-1)^d}{2^d}=
\begin{cases}
0, & \!\!\textrm{even } d\;\\
2^{1-d}, & \!\!\textrm{odd } d\;\\
\end{cases}
\end{eqnarray}

%


\begin{thebibliography}{99}
\bibitem{entanglement_review} \Name{Calabrese P., Cardy J. \and Doyon B.} (Eds.)
\textit{Entanglement entropy in extended quantum systems} (special issue), \REVIEW{J. Phys. A}{42}{2009}{500301}.

\bibitem{amico} \Name{Amico L., Fazio R., Osterloh A. \and Vedral V.} \REVIEW{Rev. Mod. Phys.}{80}{2008}{517}.

\bibitem{holzhey} \Name{Holzhey C., Larsen F. \and Wilczek F.} \REVIEW{Nucl. Phys. B}{424}{1994}{443}.

\bibitem{vidal} \Name{Vidal G., Latorre J. I., Rico E. \and Kitaev A.} \REVIEW{Phys.
Rev. Lett.}{90}{2009}{227902}.

\bibitem{Calabrese_Cardy04} \Name{Calabrese P. \and Cardy J.} \Review{J. Stat. Mech.} \Year{2004} \Page{P06002}.

\bibitem{Calabrese_Lefevre}
\Name{Calabrese P. \and Lefevre A.} \REVIEW{Phys. Rev. A}{78}{2009}{032329}.

\bibitem{area}
 \Name{Eisert J., Cramer M. \and Plenio M. B.} \REVIEW{Rev. Mod. Phys.}{82}{2010}{277}.

\bibitem{boson} \Name{Barthel T., Chung M.-C. \and Schollw\"ock U.} \REVIEW{Phys. Rev. A}{74}{2006}{022329}; \Name{Cramer M., Eisert J. \and Plenio M. B.} \REVIEW{Phys. Rev. Lett.}{98}{2007}{220603}.

\bibitem{fermion} \Name{Wolf M. M.} \REVIEW{Phys. Rev. Lett.}{96}{2006}{010404}; \Name{Gioev D.
\and Klich I.} \REVIEW{Phys. Rev. Lett.}{96}{2006}{100503}.

\bibitem{2d_Ising} \Name{Tagliacozzo L., Evenbly G. \and Vidal G.} \REVIEW{Phys. Rev. B}{80}{2009}{235127}.

\bibitem{2d_Heisenberg} \Name{Song H. F., Laflorencie N., Rachel S. \and Le Hur K.} \REVIEW{Phys. Rev. B}{83}{2011}{224410}; \Name{Kallin A. B., Hastings M. B., Melko R. G. \and Singh R. R. P.} \REVIEW{Phys. Rev. B}{84}{2011}{165134}.

\bibitem{rokhsar_kivelson} \Name{Rokhsar D. S. \and Kivelson S. A.} \REVIEW{Phys. Rev. Lett.}{61}{1988}{2376}.

\bibitem{2d_conf} \Name{Fradkin E. \and Moore J. E.} \REVIEW{Phys. Rev.
Lett.}{97}{2006}{050404}; \Name{St\'ephan J.-M., Furukawa S., Misguich G. \and Pasquier V.} \REVIEW{Phys. Rev. B}{80}{2009}{184421}; \Name{Zaletel M. P., Bardarson J. H. \and Moore J. E.}
\REVIEW{Phys. Rev. Lett.}{107}{2011}{020402}.

\bibitem{refael}
\Name{Refael G. \and Moore J. E.} \REVIEW{J. Phys. A: Math. Theor.}{42}{2009}{504010}.

\bibitem{danielreview} \Name{Fisher D. S.} \REVIEW{Phys. Rev. Lett.}{69}{1992}{534}; 
        \REVIEW{Phys. Rev. B}{51}{1995}{6411}; \REVIEW{Physica A}{263}{1999}{222}.

\bibitem{mdh}
        \Name{Ma S. K., Dasgupta C. \and Hu C.-K.} \REVIEW{Phys. Rev. Lett.}{43}{1979}{1434};
        \Name{Dasgupta C. \and Ma S. K.} \REVIEW{Phys. Rev. B}{22}{1980}{1305}.

\bibitem{im} 
For a review, see: \Name{Igl\'oi F. \and Monthus C.} \REVIEW{Physics Reports}{412}{2005}{277}.

\bibitem{refael_moore04} \Name{Refael G. \and Moore J. E.} \REVIEW{Phys. Rev. Lett.}{93}{2004}{260602}.

\bibitem{laflorencie} \Name{Laflorencie N.} \REVIEW{Phys. Rev. B}{72}{2005}{140408}.

\bibitem{igloi_lin08} \Name{Igl\'oi F. \and Lin Y.-C.} \Review{J. Stat. Mech.} \Year{2008} \Page{P06004}.

\bibitem{ladder} \Name{Kov\'acs I. A. \and Igl\'oi F.} \REVIEW{Phys. Rev. B}{80}{2009}{214416}.

\bibitem{lin07} \Name{Lin Y.-C., Igl\'oi F. \and Rieger H.} \REVIEW{Phys. Rev. Lett.}{99}{2007}{147202}.

\bibitem{yu07} \Name{Yu R., Saleur H. \and Haas S.} \REVIEW{Phys. Rev. B}{77}{2008}{140402}.

\bibitem{2dRG} \Name{Kov\'acs I. A. \and Igl\'oi F.} \REVIEW{Phys. Rev. B}{82}{2010}{054437}.

\bibitem{ddRG} \Name{Kov\'acs I. A. \and Igl\'oi F.} \REVIEW{Phys. Rev. B}{83}{2011}{174207};
\REVIEW{J. Phys. Condens. Matter}{23}{2011}{404204}.

\bibitem{perco} \Name{Kov\'acs I. A.} \REVIEW{PhD thesis}{}{2012}{}.

\bibitem{MG12} \Name{Monthus C. \and Garel Th} \REVIEW{J. Phys. A: Math. Theor.}{45}{2012}{095002};
\Review{J. Stat. Mech.} \Year{2012} \Page{P01008}.

\bibitem{ilmr08} \Name{Igl\'oi F., Lin Y.-C., Rieger H. \and Monthus C.} \REVIEW{Phys. Rev. B}{76}{2007}{064421}.

\bibitem{senthil_sachdev} \Name{Senthil T. \and Sachdev S.} \REVIEW{Phys. Rev. Lett.}{77}{1996}{5292}.

\bibitem{dynamics} \Name{Igl\'oi F., Zs. Szatm\'ari \and Lin Y.-C.} \REVIEW{Phys. Rev. B}{85}{2012}{094417}



\end{thebibliography}
\end{document}